\documentclass{webofc}
\usepackage[varg]{txfonts}   % Web of Conferences font
\usepackage{hyperref}
\usepackage{url}
\usepackage{graphicx}
\usepackage{subfig}
\hypersetup{colorlinks=true,citecolor=blue,urlcolor=blue,linkcolor=blue}
\begin{document}
\title{ML-based Adaptive Prefetching and Data Placement for US HEP Systems }

% {\let\thefootnote\relax\footnote{This project is funded by the US Department of Energy (DOE) grant with award number DESC0024648. The publisher, by accepting the article for publication, acknowledges that the U.S.
%  Government retains a non-exclusive, paid up, irrevocable, world-wide license to publish or reproduce the published
%  form of the manuscript, or allow others to do so, for U.S. Government purposes. The DOE will provide public
%  access to these results in accordance with the DOE Public Access Plan (http://energy.gov/downloads/doe-public
% access-plan).} }
{\let\thefootnote\relax\footnote{\copyright  The Authors, published by EDP Sciences. This is an open access article distributed under the terms of the Creative Commons
Attribution License 4.0 (http://creativecommons.org/licenses/by/4.0/).}}

\author{\firstname{Venkat Sai Suman} \lastname{Lamba Karanam}\inst{1}\fnsep\thanks{\email{saisuman1@acm.org}} \and
        \firstname{Sarat Sasank} \lastname{Barla}\inst{1}\fnsep\thanks{\email{sbarla2@huskers.unl.edu}} \and
        \firstname{Byrav} \lastname{Ramamurthy}\inst{1}\fnsep\thanks{\email{ramamurthy@unl.edu}}
        \and 
        \firstname{Derek} \lastname{Weitzel}\inst{1}\fnsep\thanks{\email{dweitzel@unl.edu}}   
}

\institute{School of Computing (SoC), University of Nebraska-Lincoln (UNL) }

\abstract{Although benefits from caching in US HEP are well-known, current caching strategies are not adaptive i.e they do not adapt to changing cache access patterns. Newer developements such as High Luminosity - Large Hadron Collider (HL-LHC), Deep Underground Neutrino Experiment (DUNE), a steady move toward streaming readout based Data Acquisition systems (DAQs) will increase the data production exponentially and hence burden the storage, compute \& network infrastructures. Moreover, existing caching frameworks are optimized to reduce latency, but not optimized for storage. This in combination with limited cache capacities relative to total data makes it difficult to achieve data locality. 

In this work, we present Machine Learning-aided (ML) caching strategies. Specifically, first we present a Long Short-Term Memory-based (LSTM) \textit{hourly} cache usage prediction. Second, we present an \textit{hourly file-level access prediction} model based on CatboostRegressor. To date, most ML-based cache prediction strategies in HEP have focused on \textit{daily cache usage} and limited works tackled \textit{hourly cahe usage} and even less strategies addressed \textit{hourly file-level access prediction}. \textit{File-level access prediction} allows for the design of intelligent prefetching and data placement strategies with fine-grained control. We validated our cache prediction strategies using data collected from SoCal MINI caches in August 2024. We are currently extending WRENCH simulator to reflect the US HEP exosystem at the storage, network and compute levels. We plan to deploy our cache prediction strategies into WRENCH and later perform extensive anlaysis with complex data access patterns and candidate infrastructure configurations.
}
\maketitle
\section{Introduction}
\label{sec:intro}
Caching in US HEP systems enables smooth execution of workflows by delivering the data over network to multiple jobs. Although the benefits of caching in the HEP infrastructure are well established~\cite{1}, there are existing challenges to realize the full potential of caching. Although colocating data with computation helps to increase overall efficiency, it is not always possible. In fact, the current US HEP storage infrastructure does not have enough capacity to meet the data locality for every workflow. For example, only 20 TB per cache is allocated for CMS NANOAOD origins that cumulatively host a total of approximately 100 TB. There are also a limited number of dedicated caches per site. Moreover, current caching systems are optimized for speed over capacity. All of this implies an increased data movement between caches and compute \& analyses sites. %CMS data lake proposal

Several recent advances in US HEP systems at the infrastructure and software level will enable increasingly complex experiments. These experiments will not only produce more data but also at faster rates, which needs to be analyzed quickly to make discoveries. The second phase of High Luminosity-Large Hadron Collider (HL-LHC)~\cite{HL-LHC}, expected to begin in June 2030, is expected to increase data production rates by 10X. Deep Underground Neutrino Experiment (DUNE)~\cite{DUNE} is expected to be partly operational to produce data near the end of the year 2028. Additionally, newer generations of Data Acquisition systems (DAQs) are moving toward streaming readout systems, navigating away from the traditional triggered systems~\cite{streaming, EPN2EOS}. These newer DAQ systems offer continuous \& real-time data calibration, reconstruction, and storage by offloading to remote sites results. 

The above observations imply that the available network, storage, and compute resources must be used efficiently. Traditional infrastructure upgrades are slower to implement and hence complementarily we must design intelligent strategies that reduce burden from unnecessary data movements. Current caching strategies in US HEP are not intelligent and adaptive. Machine Learning (ML) approaches help design data-driven intelligent caching strategies from analyzing cache usage logs. ML-aided intelligent caching allows for prefetching and data placement strategies that will reduce unnecessary data movements. Such systems will also improve utilization of available network, storage, and compute resources.
%%%%%
%This finding is in agreement with existing literature [4 citation from PPT], which showed that CPU efficiency and latency approximately are same between local reads at non-custodial facilities (local disk) and cache reads from custodial sites (origins and caches). On the other hand, remote non-custodial reads (i.e. remote disk reads) showed worse performance.%may be use this to support hwy caching is important??
%%%%%

%For bibliography use \cite{RefJ}, \cite{RefB}
%Don't forget to give each section, subsection, subsubsection, and paragraph a unique label (see Sect.~\ref{sec-1}).
\section{Caching in US HEP}\label{sec:2}
Caching in HEP allows for data reuse and thus reduces network usage and load on the data origins. Caching in combination with prefecthing reduces job latency and improves IO efficiency. Caching frameworks in US HEP, such as the Open Science Grid (OSG) Open Science Data Federation (OSDF) (formerly StashCache~\cite{1}) and US CMS SoCal AOD data cache, are primarily based on XCache~\cite{XCache}, also known as the XRoot disk-based file proxy cache, the primary caching service in XRootD. These caching frameworks deliver data to compute \& analyses sites from remote origins in blocks. This data is then stored on local disk via a write queue, with blocks belonging to older files deleted when the disk storage capacity is full.

\subsection{Current Caching Strategies are Optimized for Speed}%but not storage

Current caching strategies in US HEP prioritize data reuse to reduce latency. For example, OSDF is optimized to deliver the same data to multiple jobs with varying parameters. This is achieved by retaining files in popular datasets that are accessed more frequently. Although this strategy reduces latency by reducing network remote reads from origins, the limited cache capacity implies that not all popular data can be cached. So the missing dataset may still be transferred multiple times from remote sites (both caches and origins).

Difficulty in achieving data locality due to limited cache capacity will only become more difficult from newer experiments of HL-LHC and DUNE. Both experiments are expected to increase the data produced exponentially, thereby burdening the underlying storage and compute infrastructure. 

Although newer data formats were proposed, to reduce the storage sizes considerably, their adoption is currently not widespread. For example, US CMS moved from RECO to AOD, and then later to MiniAOD and NANOAOD. Although NANOAOD offers the most benefit, as of 2022 (end of LHC Run 2), only 30\% of the analyses adopted NANOAOD but are expected to increase up to 50\% by the end of Run 3~\cite{HL-LHC}. Current US ATLAS strategy is to unpin all data once a copy is created on tape~\cite{HL-LHC}. The intuition behind this is that it only retains popular data in disk and unpopular data are retained in tape (at the custodial). This allows for older \& stale data to be replaced with newer \& popular data.

In summary, existing caching strategies in US HEP are optimized to reduce latency and hence do not actively reduce the data movements to and from the caches. As a solution, cache access patterns must be leveraged to design intelligent caching strategies. Studying cache access patterns from the logs offers insights into future potential accesses. Machine Learning (ML) can be used to predict future cache accesses and help implement intelligent cache placement and prefetching, thereby reducing the number of data movements by improving the data locality adaptively.

%Cache access prediction allows intelligent caching techniques beyond blindly retaining the most recent popular files in the cache.

\subsection{Cache Access Patterns}
\label{subsec:2.1}
We analyzed the cache access patterns from logs collected from the US CMS SoCal MINI Repo during August 2024. SoCal MINI serves Tier-2 UCSD and Caltech sites. It is currently composed of approximately 15 data servers, with a majority of them residing at UCSD, two or three in Caltech, and one in an ESNet POP in Sunnyvale, California. 

\begin{figure*}[ht]
\captionsetup[subfigure]{labelformat=empty}
\centering
	\subfloat[]{\includegraphics[height=0.2555\linewidth]{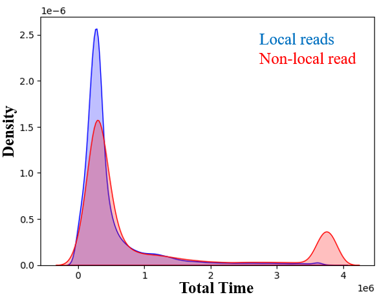}
		\label{total-read-30}
	}
	\subfloat[]{\includegraphics[height=0.245\linewidth]{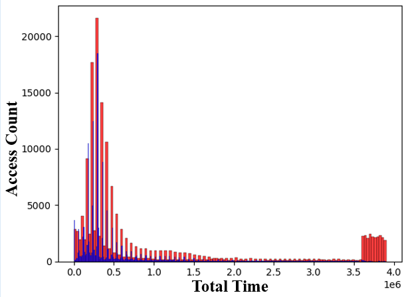}
		\label{read-vs-access-30}
	}
\caption{Kernel density of total read time (\textit{ms}) for local and non-local reads (\textit{left}) and access count vs total read time (\textit{ms}) for local and non-local reads (\textit{right)} over a 30-day period in August 2024.}
\label{fig:dist_30days}
\end{figure*}

\begin{figure*}[ht]
\captionsetup[subfigure]{labelformat=empty}
\centering
	\subfloat[]{
		\includegraphics[height=0.255\linewidth]{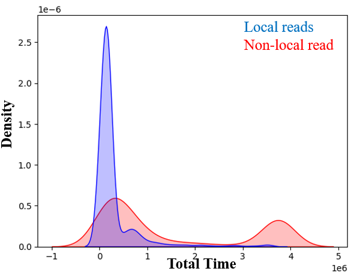}
		\label{total-read-1}
	}
	\subfloat[]{
		\includegraphics[height=0.245\linewidth]{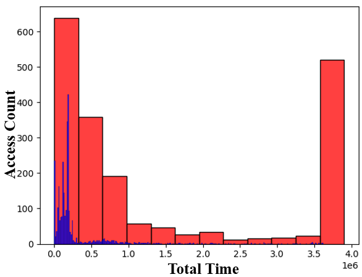}
		\label{read-vs-access-1}
	}
\caption{Kernel density of total read time (\textit{ms}) for local and non-local reads (\textit{left}) and access count vs total read time (\textit{ms}) for local and non-local reads (\textit{right)} over a one-day period  (24hrs) in August 2024.}
\label{fig:dist_1day}
\end{figure*}

\begin{figure*}[ht]
\captionsetup[subfigure]{labelformat=empty}
\centering
\subfloat[]{
		\includegraphics[height=0.25\linewidth,trim={11.25cm 6cm 10.75cm 4.5cm},clip]{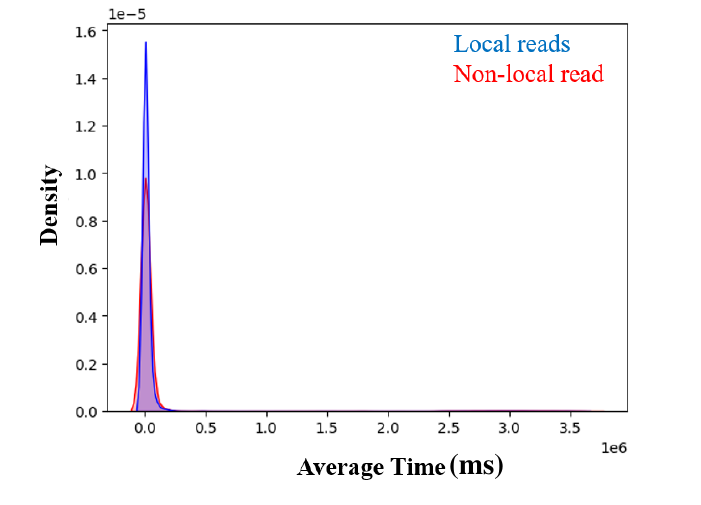}
		\label{avg-read-30}
	}
	\subfloat[]{
		\includegraphics[height=0.255\linewidth]{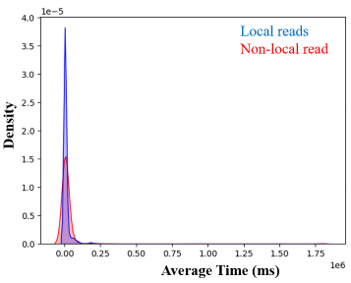}
		\label{avg-read-1}
	}
\caption{Distribution of average reads over 30 days (\textit{left}) and over one day (24 hrs) (\textit{right}).}
\label{fig:dist_avg_read_time}
\end{figure*}

As mentioned previously, limited cache capacity relative to the total data results in network data movement from remote reads for missing data. To understand the impact of non-local reads i.e. remote reads from remote caches or directly from origins, we studied the distribution of local and non-local reads at SoCal MINI. We performed preliminary analysis of the local and non-local reads. Figure~\ref{fig:dist_30days} shows the distribution of \textit{total read time} and \textit{access count vs. total read time} for local and non-local reads over a period of 30 days.  We found that non-local reads show two peaks in both \textit{left} and \textit{right} subplots. On the other hand, local reads only show one peak. The extra peak for non-local reads in both \textit{left} and \textit{right} subplots indicates that \textit{some} non-local reads take longer than others, while local reads do not indicate any such behavior.
To further investigate the validity of the assumption that \textit{some} non-local reads may be taking longer, we plotted the same but over a period of one day i.e. over 24 hours in figure~\ref{fig:dist_1day}. Similar to figure~\ref{fig:dist_30days}, \textit{left} and \textit{right} subplots in figure~\ref{fig:dist_1day} shows similar behavior with non-local reads. Our results in figures~\ref{fig:dist_30days} and~\ref{fig:dist_1day} indicate the possibility of some caches being slower than others.
However, the distribution of average read times shows that local and non-local reads follow a similar distribution (see figure~\ref{fig:dist_avg_read_time}). This indicates that the peaks observed in figures~\ref{fig:dist_30days} and~\ref{fig:dist_1day} pertain to larger files. 

Our preliminary analysis over local and non-local average read times at SoCal MINI indicate similar speed. %in cases like these, we need to reduce data movement---> cache capcaity is the bottleneck
However, we must note that although the non-local cache reads take the same time as local cache reads, there exists a portion of the dataset(s) that are accessed multiple times non-locally. This is due to the limitation in cache capacities relative to the total data. Non-local reads, even if fast, induce burden on network usage and IO bandwidth.

\section{ML-based Adaptive Caching}

%Prefetching solutions exist, but they are old. Current prefetching solution in XCache reads all blocks \textit{in order} and cannot keep up with faster compute and read rates [XCache Developments and Plans 2023].
Neural Networks such as Long Short-Term Memory (LSTM)~\cite{LSTM} and Gated Recurrent Units (GRUs)~\cite{GRU} can be used to predict cache usage into the near-future. LSTM and GRUs are exceptional for time-series forecasting and are no stranger to cache prediction in HEP [CHEP 2022 from PPT]. Previously, cache usage prediction was done on a \textit{daily} basis. However, an \textit{hourly} cache usage prediction model is more useful for HEP jobs. An \textit{hourly} prediction allows finer control in prefetching the data that may potentially be used in the future and to  to reduce the data transfer redundancy by retaining data that are likely to be accessed in the near-future. %this sentence is not complteley clear

\subsection{LSTM for Cache Usage Prediction: SoCal MINI Case Study}\label{subsec:lstm}
\begin{figure*}[ht]
\captionsetup[subfigure]{labelformat=empty}
\centering
\includegraphics[width=0.6\linewidth,trim={2cm 7cm 13.75cm 1.5cm},clip]{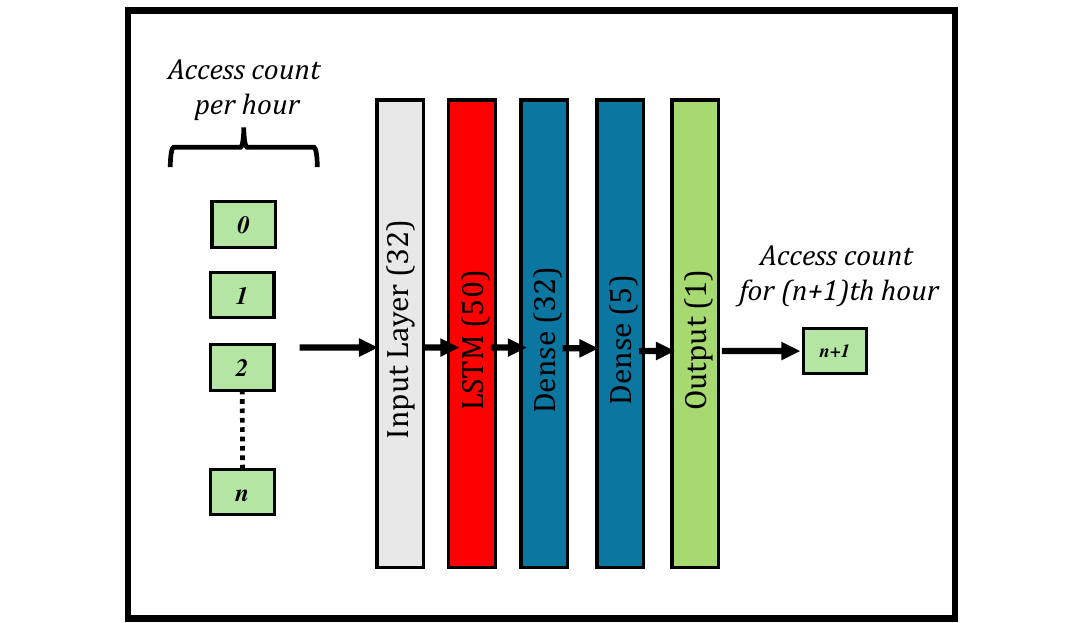}
\caption{LSTM architecture to predict cache usage for the next hour.}
\label{fig:lstm1}
\end{figure*}

Figure~\ref{fig:lstm1} shows our LSTM model architecture. The input to the model is the cache access count \textit{per hour}. The model predicts the cache access count for the \textit{next hour}. Predicting cache usage \textit{per hour} at each cache can help in several ways. First, existing data redundancy can be exploited by adaptively choosing cache sites that are predicted to be burdened. Second, data placement  strategies can be designed to reduce the burden on certain caches in the future. Third, intelligent prefetching can be designed to prefetch files or data that may be needed from a cache that is predicted to be burdened.

\begin{figure*}[ht]
\captionsetup[subfigure]{labelformat=empty}
\centering
\subfloat[]{\includegraphics[height=0.25\linewidth]{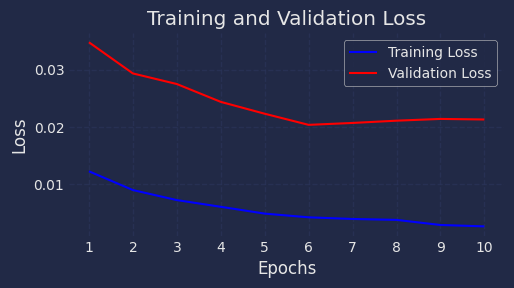}
	}
	\subfloat[]{\includegraphics[width=0.5\linewidth,trim={7cm 6cm 9cm 4.5cm},clip]{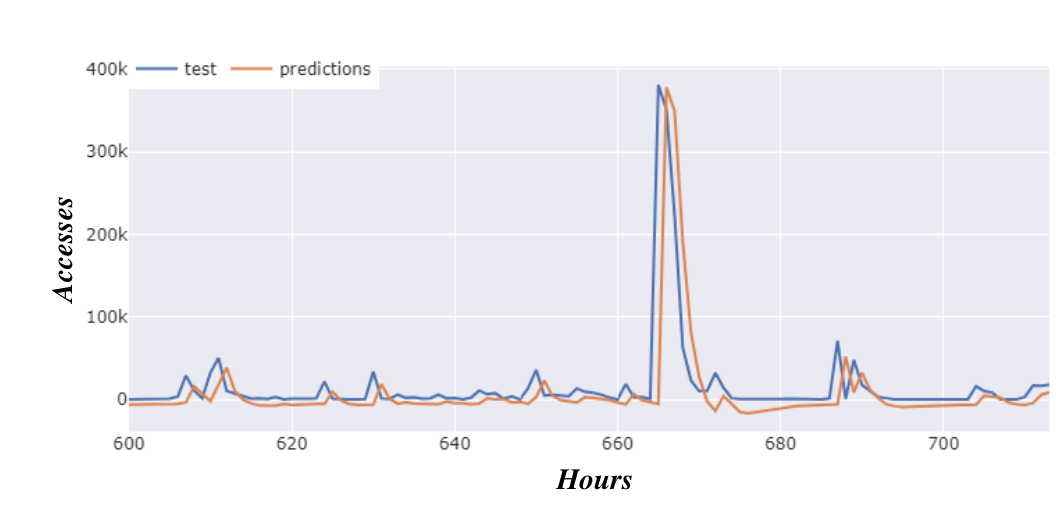}
	}
\caption{Train and validation loss (\textit{left}) and predicted vs actual \textit{hourly} cache usage (\textit{right}) for SoCal cache, which we call \textit{site1}. }
\label{fig:site1-results}
\end{figure*}

\begin{figure*}[ht]
\captionsetup[subfigure]{labelformat=empty}
\centering
\subfloat[]{\includegraphics[height=0.25\linewidth]{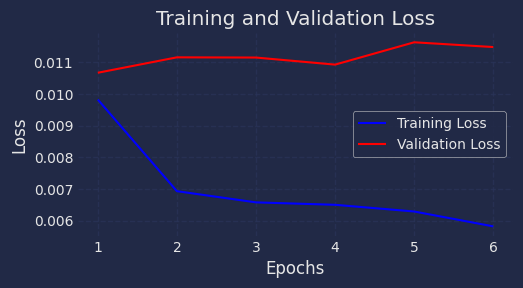}
	}
	\subfloat[]{\includegraphics[width=0.5\linewidth,trim={6cm 6.25cm 9cm 4.5cm},clip]{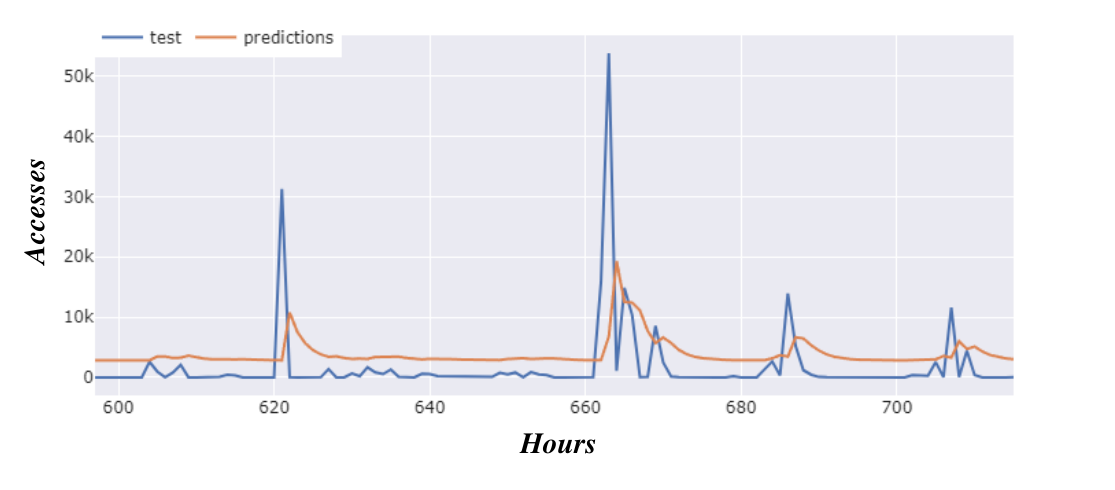}
	}
\caption{Train and validation loss (\textit{left}) and predicted vs actual \textit{hourly} cache usage (\textit{right}) for SoCal cache \textit{site2}. }
\label{fig:site2-results}
\end{figure*}

Table~\ref{tab:lstm-hyper} shows the chosen hyperparameters for the LSTM model. We found that the choice of optimizer and other batch sizes like 8, 16, 64 only affected the model convergence but not its predictions. We chose \textit{Adam} optimizer and batch size of 32 for faster model convergence. 
\begin{table}
\centering
\caption{LSTM Model Hyperparameters.}
\label{tab:lstm-hyper}      
\begin{tabular}{ll}
\hline
Name & Chosen Value  \\\hline
Optimizer & \textit{Adam} \\
Learning rate & 0.01  \\
Batch Size & 32\\
Train-Val-Test Ratio & 4:1:1  \\
Activation & \textit{ReLU}  \\
Recurrent Activation & \textit{Sigmoid}  \\
Loss & \textit{Mean Squared Error (MSE)}  \\\hline
\end{tabular}
\end{table}
Figures~\ref{fig:site1-results} and~\ref{fig:site2-results} present the training and prediction results for our LSTM model for two SoCal caches, namely, \textit{site1} and \textit{site2}. The \textit{left} subplot shows training and validation loss, and \textit{right} subplot shows the accuracy of the predicted \textit{hourly} cache usage vs actual values. The prediction results show that the model is able to correctly predict the \textit{hourly} cache usage with relatively low error. We must note that the model performs worse in predicting sudden sharp peaks in cache usage. This is evidenced by the overall a MAE of approx. 23.88\% for \textit{site1} and 150\% for \textit{site2} are skewed by the mispredictions at peak cache usage periods. 

%multi-step prediction
Beyond predicting cache usage for the \textit{next hour}, we explored prediction over \textit{several hours} into the future. This multi-step prediction i.e., predicting cache usage of more than one step (i.e. one hour) performed relatively worse when there are unexpected peaks in accesses. We do not present the results for this due to brevity.

\subsection{File-level Cache Access Prediction}\label{subsec:catboost}
%file-level prediction
While predicting overall \textit{hourly} cache usage can help design intelligent data placement and prefetching techniques, a \textit{file-level hourly access prediction} allows for intelligent file-level prefetching, data placement and even data retention strategies. We initially experimented with the same LSTM model presented in Sect.~\ref{subsec:lstm} to predict the future \textit{hourly accesses per file}. However, we found that the LSTM model underperformed in file-level predictions and also suffered from longer time to converge.
\begin{table}
\centering
\caption{CatboostRegressor Hyperparameters.}
\label{tab:catboost-hyper}      
\begin{tabular}{ll}
\hline
Name & Chosen Value  \\\hline
Learning rate & 0.01  \\
Max depth & [3, 5, 10]\\
Train-Val-Test Ratio & 4:1:1  \\
\#Estimators & [100, 500]  \\
\#Steps & 36  \\
Lags Grid & [3,5]  \\\hline
\end{tabular}
\end{table}
We designed a CatboostRegressor~\cite{catboost} to predict the \textit{hourly file-level accesses}. The hyperparameters for the CatboostRegressor are shown in Tab.~\ref{tab:catboost-hyper}. We chose a SoCal MINI cache, say \textit{site3}, comprising 2918 unique files over a period of a month. Our analysis on this dataset found that \textbf{CatboostRegressor is able to predict the \textit{next hour's access count for each file} with a \textit{Mean Absolute Error (MAE) = 1.131 and Mean Absolute Percentage Error (MAPE) = 1.0427}. To put this result in perspective, the overall \textit{average hourly file access count} for all files in\textit{site3} was 2.928.} This means that our CatboostRegressor was able to predict \textit{hourly file-level accesses} with relatively low error.  Our analysis with \textit{site1} and \textit{site2} showed very similar \textit{MAE} values. We let go of further analysis of our file-level cache access prediction approach for brevity and emphasize our results as a proof of concept. To date, most cache prediction using ML focused on predicting overall cache usage or directory-level accesses, but file-level access predictions remained to be solved. Our \textit{MAE} and \textit{MAPE} values for \textit{hourly file-level access prediction} shows that such solutions are within reach, provided enough access logs.

%\subsection{Integration into WRENCH Simulator}

\section{Conclusion and Future Work}
US HEP ecosystem is set to undergo major developments with HL-LHC, DUNE, and newer streaming readout based DAQs. Althought current caching frameworks reduce latency of reads, they incur excessive data movements due to the limited cache capacities. The presented ML-based adaptive cache prediction strategies are key to address the gaps in current caching frameworks. Our immediate future steps include hyperparameter optimization of the two cache prediction strategies and comparison against other cache datasets, such as those based on NANOAOD like the Purdue\_NANO and Nebraska\_NANO. %Concurrently, we will deploy our ML-based cache prediction approaches in WRENCH simulator to relieve the expectation of dataset availability across multiple complex scenarios. This effort will allow for performance analysis against complex data access patterns and infrastructure configurations. 

We are currently extending the WRENCH simulator~\cite{WRENCH} with the ML-aided cache prediction strategies. The simulation model underlying WRENCH has been experimentally validated~\cite{TOMACS,WRENCH-HEP}.
Simulator integration allows rapid prototyping and flexibility with several candidate storage, network \& compute configurations without testbed involvement. We integrated the ESNet and Internet2 network infrastructure capabilities into WRENCH. Next, we plan to integrate the storage/cache architecture of SoCal Repo. Following that, we will deploy the ML-aided caching strategies (see Sect.~\ref{subsec:lstm} and~\ref{subsec:catboost}) into our extended-WRENCH and perform simulations. Once we are able to validate the simulation results, we plan to test complex data and cache access patterns.
\section*{Acknowledgement}
The authors would like to thank Diego Davila from the San Diego Supercomputer Center for his support in providing the datasets used in this work. This project is funded by the US Department of Energy (DOE) grant with award number DE-SC-0024648.
Any opinions presented in this talk reflect only the opinions of the authors and not of DOE.

\end{document}